\documentclass[12pt,preprint]{aastex}

\def\msol{\hbox{\kern 0.20em $M_\odot$}}
\def\lsol{\hbox{\kern 0.20em $L_\odot$}}
\def\rsol{\hbox{\kern 0.20em $R_\odot$}}
\def\sr{\hbox{\kern 0.20em sr}}
\def\srmu{\hbox{\kern 0.20em sr$^{-1}$}}
 
\def\g{\hbox{\kern 0.20em g}}
\def\gmu{\hbox{\kern 0.20em g$^{-1}$}}
\def\kg{\hbox{\kern 0.20em kg}}
\def\pc{\hbox{\kern 0.20em pc}}
 
\def\mum{\hbox{\kern 0.20em $\mu$m}}
\def\mumd{\hbox{\kern 0.20em $\mu$m$^{-2}$}}
\def\cm{\hbox{\kern 0.20em cm}}
\def\m{\hbox{\kern 0.20em m}}
\def\km{\hbox{\kern 0.20em km}}
\def\nm{\hbox{\kern 0.20em nm}}
 
\def\s{\hbox{\kern 0.20em s}}
\def\h{\hbox{\kern 0.20em h}}
\def\sec{\hbox{\kern 0.20em sec}}
\def\min{\hbox {\kern 0.20em min}}
\def\smu{\hbox{\kern 0.20em s$^{-1}$}}
\def\smd{\hbox{\kern 0.20em s$^{-2}$}}
\def\an{\hbox{\kern 0.20em an}}
\def\anmu{\hbox{\kern 0.20em an$^{-1}$}}
\def\deg{\hbox{\kern 0.20em $^{\rm o}$}}
\def\yr{\hbox{\kern 0.20em yr}}
\def\yrmu{\hbox{\kern 0.20em yr$^{-1}$}}
\def\Myr{\hbox{\kern 0.20em Myr}}
\def\Mymu{\hbox{\kern 0.20em Myr$^{-1}$}}
\def\K{\hbox{\kern 0.20em K}}
\def\pcmu{\hbox{\kern 0.20em pc$^{-1}$}}
\def\pcmd{\hbox{\kern 0.20em pc$^{-2}$}}
\def\pcmt{\hbox{\kern 0.20em pc$^{-3}$}}
\def\kms{\hbox{\kern 0.20em km\kern 0.20em s$^{-1}$}}
\def\kmpd{\hbox{\kern 0.20em km$^{2}$}}
\def\kpc{\hbox{\kern 0.20em kpc}}
\def\cms{\hbox{\kern 0.20em cm\kern 0.20em s$^{-1}$}}
\def\erg{\hbox{\kern 0.20em erg}}
\def\ergs{\hbox{\kern 0.20em erg}}
\def\cmpd{\hbox{\kern 0.20em cm$^2$}}
\def\cmmd{\hbox{\kern 0.20em cm$^{-2}$}}
\def\cmms{\hbox{\kern 0.20em cm$^{-6}$}}
\def\cmpt{\hbox{\kern 0.20em cm$^3$}}
\def\cmmt{\hbox{\kern 0.20em cm$^{-3}$}}
\def\mpd{\hbox{\kern 0.20em m$^2$}}
\def\mmd{\hbox{\kern 0.20em m$^{-2}$}}
\def\mpt{\hbox{\kern 0.20em m$^3$}}
\def\mmt{\hbox{\kern 0.20em m$^{-3}$}}
\def\mujy{\hbox{\kern 0.20em $\mu$Jy}}
\def\mjy{\hbox{\kern 0.20em mJy}}
\def\Mj{\hbox{\kern 0.20em MJy}}
\def\jy{\hbox{\kern 0.20em Jy}}
\def\ghz{\hbox{\kern 0.20em GHz}}
\def\srmd{\hbox{\kern 0.20em sr$^{-1}$}}
\def \kms{km~$\rm{s}^{-1}$}

\def \mum{$\mu$m}

\def\G{\hbox{\kern 0.20em G}}

\def\h13cop{\hbox{H$^{13}$CO$^{+}$}}

\def\S+{\hbox{S{\small II}}}

\slugcomment{Accepted for publication, ApJ, on March 2011}

\shorttitle{Spitzer Observations of HH~111}
\shortauthors{Noriega-Crespo et al.}
\def \mum{$\mu$m}

\begin{document}

\title{The precession of the HH~111 flow in the infrared}

\author{Noriega-Crespo, A.\altaffilmark{1},
Raga, A. C. \altaffilmark{2},
Lora, V.\altaffilmark{3},
Stapelfeldt, K. R.\altaffilmark{4}
and
Carey, S. J.\altaffilmark{1}}

\altaffiltext{1}{SPITZER Science Center, California Institute of 
Technology,CA 91125  USA}
\altaffiltext{2}{Instituto de Ciencias Nucleares, Universidad Nacional 
Aut\'onoma de M\'exico, Ap. 70-543, 04510 D.F., M\'exico}
\altaffiltext{3}{Astronomisches Rechen-Institut Zentrum f\"ur
Astronomie der Universit\"at Heidelberg, M\"onchhofstr. 12-14
69120 Heidelberg, Germany }
\altaffiltext{4}{Jet propulsion Laboratory, California Institute of 
Technology,MS 183-900, 4800 Oak Grove Drive, Pasadena, CA 91109, USA}

\begin{abstract}
We present Spitzer IRAC images of the HH~111 outflow, that
show a wealth of condensations/knots in both jet and counterjet.
Studying the positional distribution of these knots, we find
very suggestive evidence of a mirror symmetric pattern
in the jet/counterjet flow. We model this pattern as the result
of an orbital motion of the jet source around a binary companion.
From a fit of an analytic, ballistic model to the observed
path of the HH~111 system, we find that the motion in a binary with two
$\sim 1$\msol\ stars (one of them being the HH~111 source),
in a circular orbit with a separation of $\sim 186$~AU would produce
the mirror symmetric pattern seen in the outflow.
\end{abstract}


\section{Introduction}

Over the past two decades considerable progress has been made
in understanding the formation of stellar jets and their influence on 
the surrounding interstellar medium (see, e. g., Bally, Reipurth \& Davis 2007;
Arce et al. 2007). Nevertheless, some fundamental questions remain unanswered,
mostly related with the nature of the acceleration and collimation of the jets
themselves, the properties of the 'launch region' and how both of
these processes can be influenced by nearby objects, since
many outflows arise from binary or multiple young stellar systems.

Since the discovery of HH 111 (Reipurth 1989) 
using standard narrow band imaging techniques to map the outflow
in H$\alpha$ and [S~II]$\lambda\lambda$ 6717/6731, HH 111 has become an icon
of highly collimated stellar outflows, and has been the subject of
a number of studies across the entire wavelength spectrum to determine 
the proper motions of their 'knots' (Hartigan et al. 2001; 
Coppin, Davis \& Micono 1998),
kinematics (Reipurth, Raga \& Heathcote 1992; Raga et al. 2002),
ionization structure (Reipurth 1989; Davis, Hodapp \& Desoches 2001; 
Nisini et al. 2002), circumstellar environment (Stapelfeldt \& Scoville 1993; 
Yang et al. 2007), shock properties (Noriega-Crespo, Garnavich \& Raga 1993; 
Morse et al. 1993), molecular flow morphology (Cernicharo \& Reipurth 1996; 
Lefloch et al. 2007), and ejection history (Masciadri et al. 2002; 
Raga et al. 2002), among others. The original optical images were of the 
West lobe of HH~111, and the Eastern counterjet was later identified in the
near infrared (NIR) (Gredel \& Reipurth 1994; Davis, Mundt \& Eisl\"offel 
1994).

HH~111 lies in the L1617 cloud in Orion at a distance of $\sim 417$pc and is
oriented almost E-W (PA$\sim$97.5\arcdeg) with a length of $\sim$ 11\arcmin~
in our IRAC images (see Fig 1). HH~111 is part of a quadrupolar flow 
(Reipurth et al. 1999; Rodr\'\i guez et al. 2008), with HH~121 arising nearly 
perpendicular (PA$\sim$27.3\arcdeg) to the HH~111 flow, suggesting that 
the VLA1 source is part of a close binary system.
Reipurth et al. (1999) proposed that this binary could have a separation of
$\sim$ 0\arcsec.1~($\sim$ 50 AU). Rodr\'\i guez et al. (2008) detect two
elongated structures with a separation of $\sim$ 0\arcsec.036~($\sim$ 15 AU),
which could correspond to the previously proposed binary.

Finally, we should note that HH 111 is the central region of a ``giant HH
jet'', with two lobes extending out to $\approx 3.5$~pc from the
outflow source (Reipurth et al. 1997). The far ends of these lobes
(HH 311 and HH 113, see Reipurth et al. 1997)
show a point-symmetric deviation from the direction of the inner
HH~111 outflow of $\approx 4^\circ$. This change in direction
between the inner and outer region of the HH~111 outflow is evidence
for a precession of the outflow axis, with a period $\tau_p\approx
3.5$~pc$/150$~km~s$^{-1}\approx 23000$~yr (considering the proper
motion of $\sim 150$~km~s$^{-1}$ measured by Reipurth et al. 1997 for
HH 311 and 113).

Recently, we have shown (Raga et al. 2011) that by using observations 
from the Spitzer Space Telescope ({\it Spitzer}) (Werner et al. 2004) 
in the mid-infrared, where the extinction is considerably lower than 
in the optical and near infrared, one can use the jet/counterjet 
symmetry to set some of the strongest constraints yet on the size of 
the jet formation region. 
In this study, we take advantage once again of the superb sensitivity
and angular resolution of the Infrared Array Camera (IRAC) (Fazio et al. 2004) 
on board of {\it Spitzer} to study the positional distribution of the ``knots''
along the Herbig-Haro (HH) 111 flow, finding mirror symmetric deviations
from the jet/counterjet axis. This is the first time that such a mirror
symmetric pattern is seen in an HH outflow, which can be interpreted
in terms of the presence of a binary outflow source.

\section{Data}

The observations of HH~111 are part of our original {\it Spitzer}
General Observer (GO) program 3315 (PI Noriega-Crespo) obtained with IRAC
and the infrared photometer MIPS (Rieke et al. 2004)
in March 28, 2005. The data have been downloaded from the Spitzer Legacy
Archive and the quality of the final images (Post Basic Calibrated 
Data or Post-BCD; S18.7 products) is outstanding, so that no further 
processing was required. The HH~111 outflow is part of a larger mosaic of
the L1640 region obtained with IRAC in its four channels
(1, 2, 3, 4) = (3.6, 4.5, 5.8 \& 8.0 \mum) covering a FOV of $\sim 19\times 
23$\arcmin~ (the result of a 5$\times$4 array map with a 260\arcsec~stepsize) 
and with a total integration time per pixel of 360 sec, using 30 sec High Dynamic 
Range (HDR) exposures. The final images are sampled with 0.6\arcsec~per pixel,
nearly one third of the standard $\sim 2$\arcsec~IRAC angular resolution.

The IRAC images (Fig 1.), in comparison with the published NIR images (e.g. 
Davis, Mundt \& Eisl\"offel 1994; Fig 5), display a similar morphology, but 
given their higher signal-to-noise and lower extinction permit us to define 
better the knots and identify a couple more on each side close to the VLA1 
source. Like in other protostellar outflows observed with Spitzer 
(e.g., Noriega-Crespo et al. 2004; Looney, Tobin \& Kwon 2007; Tobin et al. 
2007; Ybarra \& Lada 2009; Raga et al. 2011) HH~111 is brighter in 
channel 2, although it is well detected in all 4 IRAC channels. 
The emission is most likely dominated by the H$_2$ pure rotational 
lines (De Buizer \& Vacca 2010).

\begin{figure*}
\centerline{\hbox{
\includegraphics[width=440pt,height=200pt,angle=0]{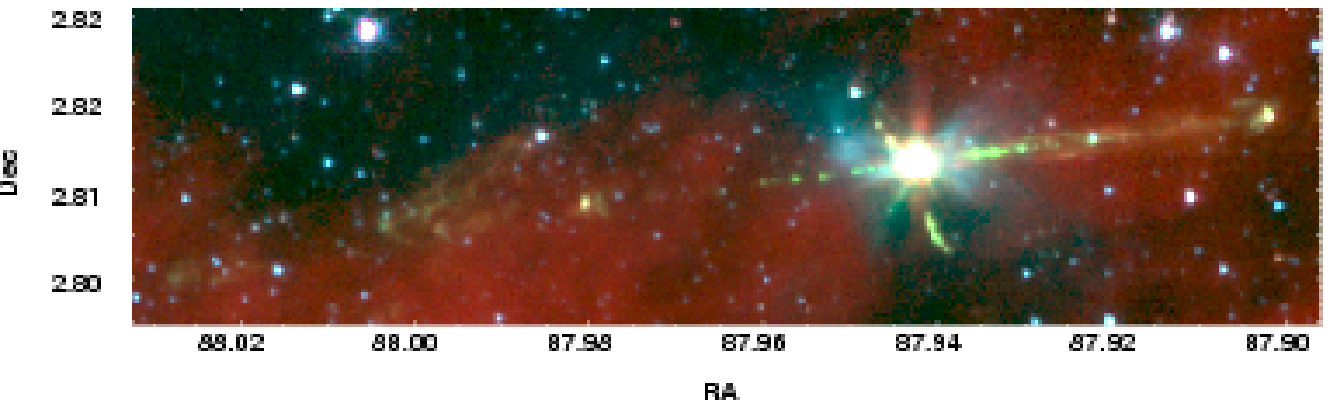}}}
\caption{The HH~111 flow in the mid-IR using the 3.6 (blue),
4.5 (green) and 8\mum~(red) IRAC channels. Notice the scattered light (blue)
near the source and the nearly perpendicular bipolar outflow HH~121.}
\label{fig1}
\end{figure*}

In the following we describe the emission peaks as condensations and/or knots.
It is clear from the HST images at high angular resolution that many of these
structures are bowshock-like (working surfaces) and that they can overlap
one another. We use the intensity peaks to define the HH 111 flow.
The measurements of the positions of the condensations/knots were carried out
using the IRAF {\it imexam} routine interactively on the IRAC channel 2 image. 
For compact ``knots'' a two-dimensional gaussian was used for determining
their centroids; for the more diffuse structures, we use the position of the
pixel with the peak emission. The uncertainty of the centroids ranges 
from 0.1 to 0.3 of a fraction of a pixel for the bright and fainter knots, 
respectively.
Since the ``knots'' are compact in most cases, we measured the flux densities using 
a circular aperture and an aperture correction. For the compact ``knots'' we 
used a  2.4\arcsec~radius for the photometric measurements with a 1.221 
aperture correction; the background was measured off the jet/counterjet flow axis.
For 7 ``knots'' that were obviously more 
extended or irregular than this aperture ($6.23\times 10^{-10}$ steradian area)
a comparable polygonal area was used without an aperture correction. Some of 
these ``knots'' do have a low surface brightness, but it was still possible 
to measure the flux density and centroid.

In Table 1 we give the relative positions in $X$ (along the outflow axis) 
and $Y$ (perpendicular to the outflow axis) with respect to the central 
source, VLA1 at RA=05:51:46.25 and Dec=02:48:29.5 (Reipurth et al. 1999) 
of the``knots'' seen along the jet and the counterjet. Table 1 contains also 
their RA and Dec coordinates, flux density and uncertainty (both in mJy).

\begin{figure}[ht]
\centerline{\hbox{
\includegraphics[width=290pt,height=500pt,angle=-90]{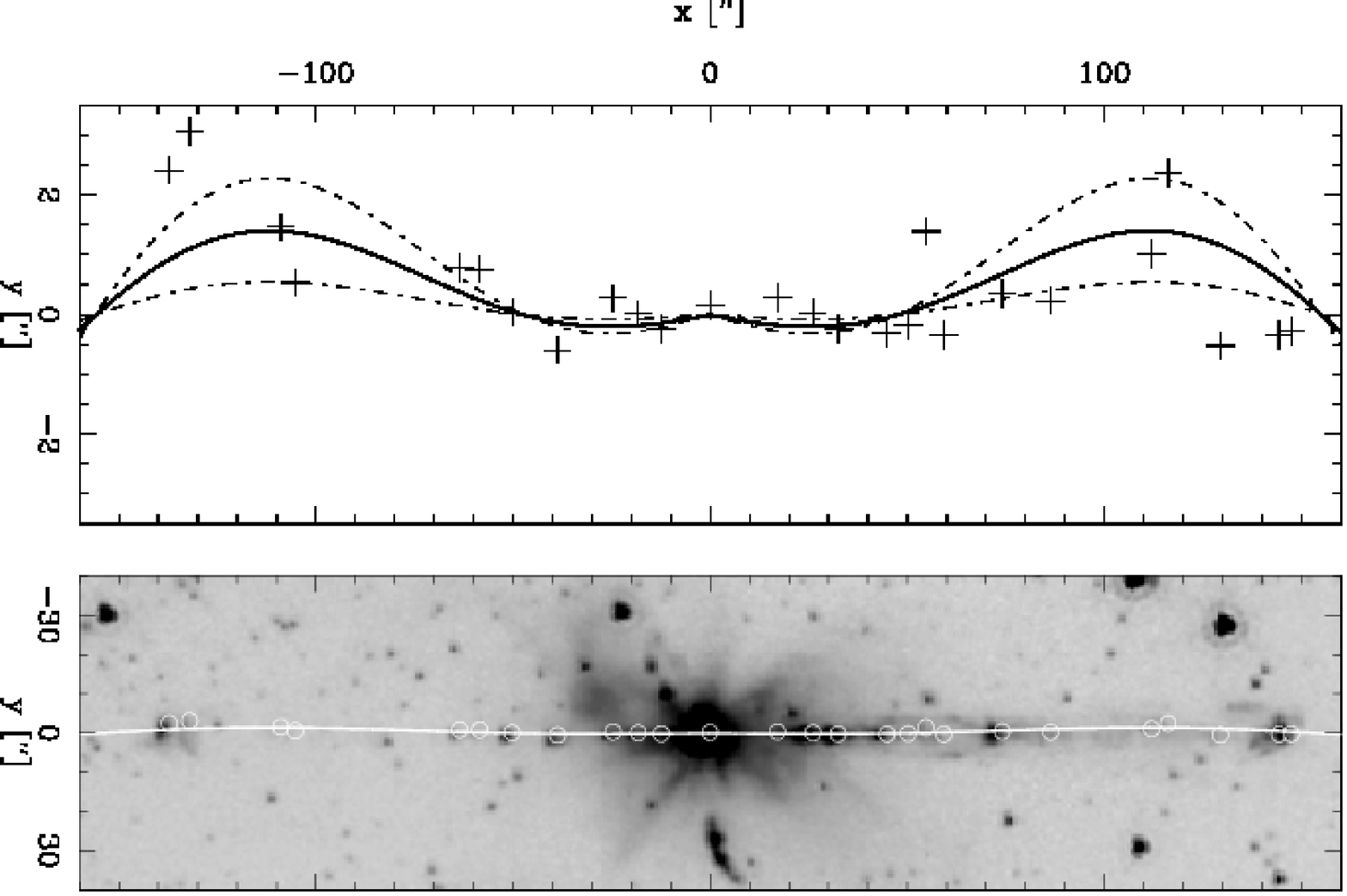}}}
\caption{The right panel shows the central region of HH~111
at 4.5\mum~rotated so that
its flow axis is vertical, and with 
the positions of the knots are shown with the white circles.
The left panel shows the same 
knots positions (shown with crosses)
with an expanded $Y$-axis (perpendicular to the outflow 
direction). The solid curve corresponds to a $\kappa=0.013$ ballistic jet
(best fit model), and the dashed
lines correspond to $\kappa=0.005$ and 0.021 (the three models
have $L=216''$ and $\phi=101^\circ$, see the text). The
jet path predicted from the best fit model is also shown
on the right hand side plot (white curve).}
\label{fig2}
\end{figure}

\section{The mirror symmetry of the HH 111 jet/counterjet system}

Figure 2 shows the inner region of the HH~111 outflow (top) with the 
measured $(X,Y)$ positions of the knots with respect to the VLA1 source.
The bottom plot shows the knot positions with an expanded $Y$-coordinate.
It is clear that both the jet and the counterjet show an excursion in the
$-Y$-direction at a distance of $\approx 40''$ from the source, and
an excursion in the $+Y$-direction at $\approx 115''$ from
the source. These excursions could be part of a mirror
symmetric spiral shape with
a step of $\sim 2\times (115''-40'')=150''$.

These mirror-symmetric excursions of the jet/counterjet system could
possibly be the effect of an orbital motion of the outflow source.
Ballistic flows from sources in circular and elliptical orbits
were described by Masciadri \& Raga (2002) and Gonz\'alez \& Raga
(2004), respectively.

For a ballistic flow, the path of the
beam of a constant velocity jet ejected parallel
to the orbital axis (of an outflow source in a circular orbit) is
given by~:
\begin{equation}
y=\kappa x\sin\left(\frac{2\pi}{\tau_ov_j}x-\psi\right)\,,
\label{y}
\end{equation}
\begin{equation}
z=\kappa x\cos\left(\frac{2\pi}{\tau_ov_j}x-\psi\right)\,,
\label{z}
\end{equation}
where $x$ is the axial coordinate, $(y,z)$ are axes parallel to the orbital
plane, $\psi$ is the orbital phase,
$\tau_o$ is the orbital period and $\kappa=v_o/v_j$ is the ratio
between the orbital and the jet velocities. As can be seen from
a comparison of Equations (\ref{y}-\ref{z}) with the results
presented by Masciadri \& Raga (2002), in these equations we have considered
that the orbital radius is negligibly small compared to $y$ and $z$.

From Equations (\ref{y}-\ref{z}), we see that the
spiral described by the jet beam has a step
\begin{equation}
L={v_j}{\tau_o}\,.
\label{l}
\end{equation}

From Equation (\ref{y}), we calculate the predicted locus
of the jet/counterjet structure, and project it onto the plane of
the sky, assuming an angle of $10^\circ$ between the orbital
axis and the plane of the sky (Reipurth et al. 1992), and that the
$y$-axis is parallel to the plane of the sky). We then carry out a least
squares fit in order to obtain the values of the model parameters
that best fit the HH 111 jet/counterjet locus within $150''$
from the outflow source. The results of the fit
are shown in the left panel of Figure 2.

The following parameters are obtained from the fit~:
\begin{enumerate}
\item an orbital phase $\phi=(101\pm 38)^\circ$. This result
implies that the line joining the two stars of the binary
outflow source currently lies close to the line of sight,
\item a spiral step $L=(216\pm 39)''$. For a distance of
414~pc to HH 111 this corresponds to a distance of
$(1.34\pm 0.24)\times 10^{18}$~cm. Also, from the proper motions measured by
Hartigan et al. (2001), we see that the knots at $\sim 40''$ along the jet
(corresponding to the first sideways excursion in the jet path, see
Figure 2) have a spatial velocity $v_j\approx 240$~km~s$^{-1}$. From
Equation (\ref{l}) we then obtain an orbital period
$\tau_o=1801\pm 317$~yr,
\item a $\kappa=0.013\pm 0.008$ (see equations \ref{y}-\ref{z}). For
a jet velocity of 240~km~s$^{-1}$ (see above), this implies an
orbital velocity $v_o=3.1\pm 1.9$~km~s$^{-1}$.
\end{enumerate}

We now use the relation (valid for a circular orbit)
between the mass of the jet source $M_1$,
the mass of the companion $M_2=\alpha M_1$, the orbital period
$\tau_o$ and the orbital radius (of the jet source) $r_o$~:
\begin{equation}
\frac{\alpha^3M_1}{(1+\alpha)^2}=\frac{\tau_o v_o^3}{2\pi G}=
(0.10,2.02,6.97)\,{\rm M_\odot}\,,
\label{m}
\end{equation}
where the three values in the third term correspond to
$(\kappa,L)=(0.005,177'')$ (the lower limits for $\kappa$
and $L$, see above), $(\kappa,L)=(0.013,216'')$ (the values
that result in the lowest $\Xi^2$) and $(\kappa,L)=(0.021,255'')$
(the upper limits).

Each of the three values of the third term of Equation (\ref{m}) corresponds
to a curve in the $M_2$ vs. $M_1$ plane, which are shown
in Figure 3. Together with
the $M_2<M_1$ condition (i.e. assuming that the source
of the strong HH~111 outflow is the more massive star of the binary),
we then obtain an allowed region for the masses of the HH~111
source binary.

From Figure 3, we see that the observed shape of the HH~111 jet/counterjet
system is consistent with a binary with two 
$\sim 1$M$_\odot$ stars. Such a mass is consistent with the masses of
T Tauri stars.

If we consider $M_1=M_2=M\approx 1$~M$_\odot$
(consistent with the curved shape of the HH 111 system, see Figures
2 and 3) and $\tau_o=1800$~yr, from the equations of a circular orbit
we then obtain a separation $2r_o\approx 186$~AU between the two stars
in the binary. Given that the orbial phase derived from the fit
implies that the stars in the binary are closely aligned with
the line of sight, the projected separation between the stars
would be substantially smaller than 186~AU.

\begin{figure}[!ht]
\centerline{\hbox{
\includegraphics[width=230pt,height=230pt,angle=0]{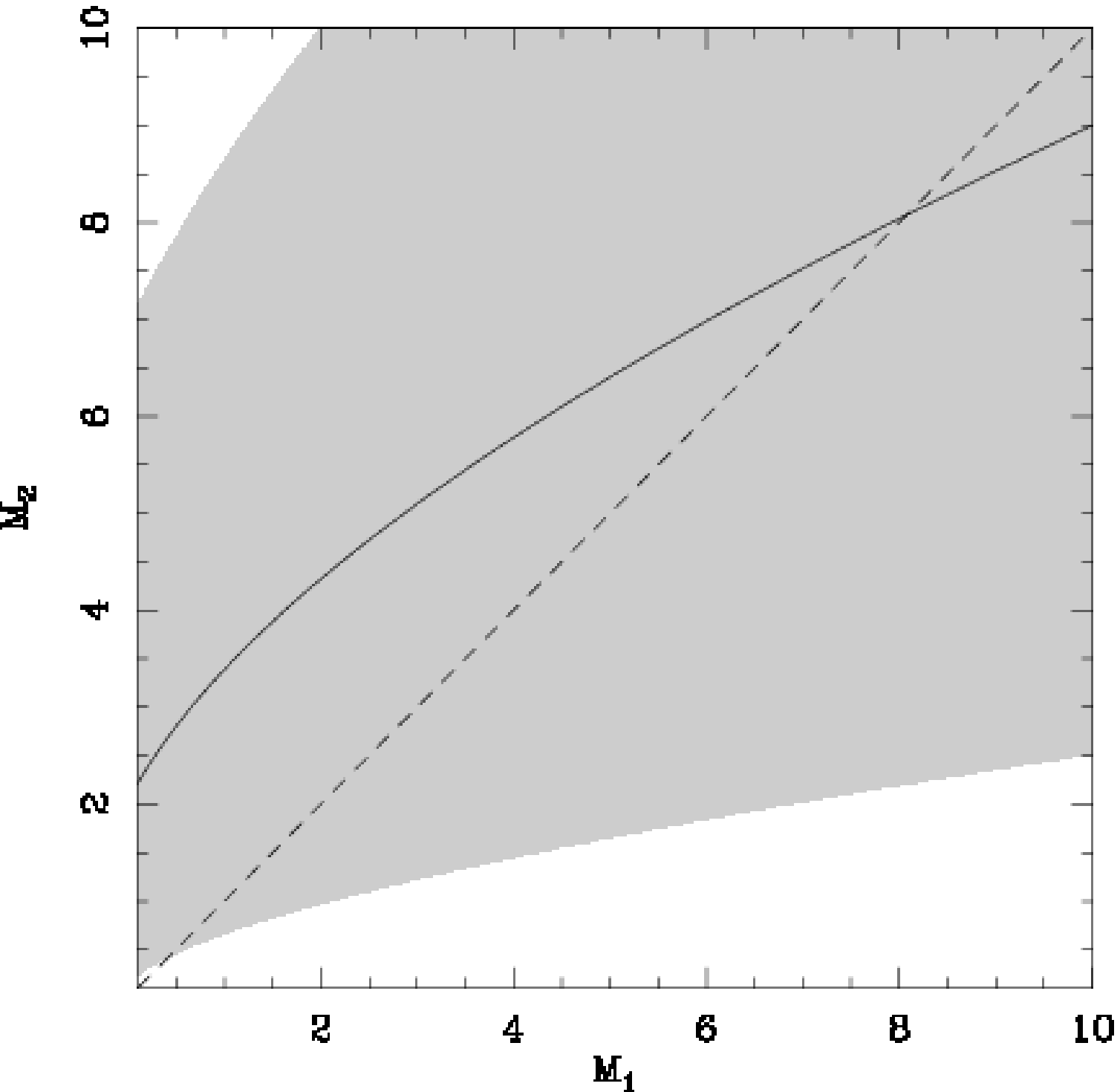}}}
\caption{Parameter space for the primary and secondary masses
of the binary system ($M_1$ and $M_2$, in units of $M_\odot$)
derived from the curved shape of the HH 111
jet/counterjet system. The two curves represent the limits
for the allowed masses derived from fitting the ballistic, orbiting
source jet model to the observed trajectory (see equation \ref{m}).
The shaded region shows the parameter space
allowed by the fit. The solid curve is deduced from
the parameters of the best fit. If one assumes that $M_2<M_1$
(i. e., that the source of the HH 111 outflow
is the more massive star in the binary), the masses of the
binary should lie beneath the $M_2=M_1$ line (the straight
diagonal cutting across the plot).}
\label{fig3}
\end{figure}

\section{Summary and conclusions}

We have presented new IRAC images of the HH~111 outflow. These images
show chains of aligned knots along the jet and the counterjet. We
have carried out astrometric measurements of the positions of these
knots, and have also determined their fluxes in the IRAC 2 (4.5 \mum)
channel.

The jet/counterjet knots within $\sim 2'$ from the outflow source
show mirror symmetric deviations from the outflow axis, which
can be interpreted as the effect of an orbital motion
of the outflow source. If we fit the observed shape of the projected
path of HH~111 with a ballistic jet model (Masciadri \& Raga 2002),
we find that the observations imply the existence of a binary
with a $\tau_o = (1801\pm 317)$~yr period, and an orbital
velocity $v_o = (3.1\pm 1.9)$~km~s$^{-1}$. This range of
periods and orbital velocities would be consistent with
a binary formed by two $\sim 1$~M$_\odot$ stars, with a separation
of $\sim 186$~AU. The model fit also implies that the stars
are currently quite closely aligned with the line of the sight,
so that this separation might be consistent with the considerably
lower projected separation implied by the observations of
Rodr\'\i guez et al. (2008).

Interestingly, the orbital period that we determine ($\tau_o\approx
1800$~yr, see above) is $\sim 13$ times smaller than the
precession period $\tau_p$ implied by the point-symmetrical ``HH 111 giant
jet'' (see Reipurth et al. 1997 and the discussion at the end of
section 1). The resulting $\tau_p/\tau_o\sim 13$ ratio is
consistent with the predictions from the precessing accretion
disk models of Terquem et al. (1999).

Raga et al. (2009) pointed out that the models of Terquem et al.
(1999) implied that outflows from binary sources would have mirror
symmetries close to the outflow source (due to the orbital motion) and point
symmetries at larger distances (if the ejection direction follows
the disk precession). Our new observations show that the HH~111 jet/counterjet
system has mirror symmetry within $\sim 2'$ from the source. This result,
combined with the point symmetry of the HH~111 giant jet (Reipurth et al.
1997) is the first evidence ever of an astrophysical jet with the
predicted flip in types of symmetry between regions close to and far
away from the outflow source.

It is evident that the mirror symmetric, side-to-side excursions
of the HH 111 jet/counterjet are a quite subtle effect. If our
interpretation of these deviations were incorrect,
we would not expect that a comparison
with a model of an outflow from an orbiting source would give
us reasonable masses for the binary components. However, we find
that a binary with two $\sim 1 M_\odot$ stars and a separation of
$\sim 200$~AU produces the observed spiral shape
of HH 111. This completely reasonable
result is a strong indication that we
might indeed be interpreting the data correctly (see also
the case for HH 211 by Lee et al. 2010).

\acknowledgements
This work is based in part on observations made with the {\it Spitzer Space 
Telescope} which is operated by the Jet Propulsion Laboratory, California 
Institute of Technology under NASA contract 1407.
The work of AR and VL was supported by the CONACyT grants 61547,
101356 and 101975.

\clearpage

\begin{deluxetable}{rrrrrr}
\tablecaption{CounterJet (East) \& Jet (West) Positions \& Flux Densities\label{tbl:EWJet}}
\tablewidth{0pt}
\tablecolumns{6}
\tablehead{\colhead{$X_{off}$(\arcsec)}  & \colhead{$Y_{off}$(\arcsec)}  & 
\colhead{RA(\arcdeg)} & \colhead{Dec(\arcdeg)}  & \colhead{F(mJy)} & \colhead{Unc(mJy)}}
\startdata
      0.00 &      0.00 &    87.9426 &    2.8084 &    123.89 &     12.39\\
    -12.39 &     -0.40 &    87.9460 &    2.8078 &      0.69 &      0.07\\
    -18.25 &     -0.14 &    87.9476 &    2.8076 &      0.46 &      0.05\\
    -24.67 &      0.12 &    87.9494 &    2.8075 &      0.32 &      0.03\\
    -38.78 &     -0.75 &    87.9533 &    2.8067 &      0.18 &      0.02\\
    -50.28 &     -0.09 &    87.9564 &    2.8065 &      0.21 &      0.02\\
    -58.53 &      0.60 &    87.9587 &    2.8064 &      0.06 &      0.01\\
    -63.61 &      0.63 &    87.9601 &    2.8062 &      0.13 &      0.01\\
   -105.40 &      0.39 &    87.9716 &    2.8045 &      0.02 &      0.01\\
   -109.17 &      1.32 &    87.9727 &    2.8047 &      0.03 &      0.01\\
   -132.28 &      2.92 &    87.9791 &    2.8042 &      0.37\tablenotemark{a} & 0.04\\
   -137.47 &      2.27 &    87.9805 &    2.8039 &      0.91\tablenotemark{b} & 0.09\\
   -199.16 &     -1.37 &    87.9973 &    2.8006 &      0.22\tablenotemark{c} & 0.02\\
   -202.54 &      8.10 &    87.9986 &    2.8031 &      0.11 &      0.01\\
   -208.84 &      8.47 &    88.0004 &    2.8029 &      0.11 &      0.01\\
   -219.96 &      8.22 &    88.0034 &    2.8024 &      0.19 &      0.02\\
   -221.45 &      3.17 &    88.0036 &    2.8010 &      0.21 &      0.02\\
   -236.14 &     -5.11 &    88.0074 &    2.7982 &      0.05 &      0.01\\
   -293.33 &     -9.45 &    88.0230 &    2.7948 &      0.06 &      0.01\\
   -309.59 &     -6.21 &    88.0275 &    2.7951 &      0.09 &      0.01\\
\tablebreak
      0.00 &      0.00 &    87.9426 &    2.8084 &    123.89 &     12.39\\
     17.12 &      0.14 &    87.9379 &    2.8090 &      0.62 &      0.06\\
     26.18 &     -0.13 &    87.9354 &    2.8093 &      2.82 &      0.28\\
     32.60 &     -0.23 &    87.9336 &    2.8095 &      2.50 &      0.25\\
     44.99 &     -0.45 &    87.9302 &    2.8099 &      0.67 &      0.07\\
     50.34 &     -0.33 &    87.9288 &    2.8101 &      0.20 &      0.02\\
     54.88 &      1.25 &    87.9276 &    2.8107 &      0.18 &      0.02\\
     59.41 &     -0.50 &    87.9263 &    2.8104 &      0.16 &      0.02\\
     74.28 &      0.20 &    87.9222 &    2.8112 &      1.61\tablenotemark{d} &  0.16\\
     86.57 &      0.08 &    87.9188 &    2.8116 &      0.14 &      0.01\\
    112.24 &      0.86 &    87.9118 &    2.8128 &      0.10 &      0.01\\
    116.44 &      2.22 &    87.9107 &    2.8133 &      0.10 &      0.01\\
    129.70 &     -0.68 &    87.9069 &    2.8130 &      0.09 &      0.01\\
    144.64 &     -0.49 &    87.9028 &    2.8136 &      0.71\tablenotemark{e} &  0.07\\
    147.73 &     -0.43 &    87.9020 &    2.8137 &      1.77 &  0.18\\
    167.50 &     -2.07 &    87.8965 &    2.8140 &      0.07\tablenotemark{f} &  0.01\\
    226.05 &     -5.10 &    87.8803 &    2.8154 &      0.04\tablenotemark{g} &  0.01\\
\enddata
{\baselineskip=0pt
\tablenotetext{a}{Over a $1.058\times 10^{-10}$ steradian area}
\tablenotetext{b}{Over a $3.976\times 10^{-10}$ steradian area}
\tablenotetext{c}{Over a $6.431\times 10^{-10}$ steradian area}
\tablenotetext{d}{Over a $8.292\times 10^{-10}$ steradian area}
\tablenotetext{e}{Over a $6.685\times 10^{-10}$ steradian area} 
\tablenotetext{f}{Over a $7.615\times 10^{-10}$ steradian area}
\tablenotetext{g}{Over a $8.377\times 10^{-10}$ steradian area}
}
\end{deluxetable}


\begin{references}{}

\reference{arce07} Arce, H.G. et al. 2007 in ``Protostars and Planets V'' 
B. Reipurth, D. Jewitt \& K. Keil eds. University of Arizona Press, p. 245
\reference{bally07} Bally, J., Reipurth, J., \& Davis, C.J. 2007 
in ``Protostars and Planets V'', B. Reipurth, D. Jewitt \& K. Keil (eds).
 University of Arizona Press, p. 215
\reference{pepe96}
Cernicharo, J. \& Reipurth, B., 1996, ApJ, 460, 57
\reference{coppin98}
Coppin, K. E. K., Davis, C. J. \& Micono, M. 1998, MNRAS, 301, 10
\reference{xris94}
Davis, C. J., Mundt, R., \& Eisl\"offel, J. 1994, ApJ, 437, 55
\reference{deb10}
De Buizer, J. M. \& Vacca, W. D. 2010, AJ, 140, 196
\reference{fas04}
Fazio, G. et al. 2004, ApJS, 154, 10
\reference{gon04}
Gonz\'alez, R. F., Raga, A. C., 2004, RMxAA, 40, 61
\reference{gredel94}
Gredel, R. \& Reipurth, B. 1994, A\&A, 289, 19
\reference{pat01}
Hartigan, P. et al. 2001, ApJ, 559, L157
\reference{lee10}
Lee, Ch.-F. et al. 2010, ApJ, 713, 731
\reference{lefloch07}
Lefloch, B. et al. 2007, ApJ, 658, 498
\reference{lon07}
Looney, L. W., Tobin,J.J., Kwon, W. 2007, ApJ, 670, 131
\reference{elena02}
Masciadri, E. et al. 2002, ApJ, 573, 260
\reference{elena02b}
Masciadri, E., Raga, A. C., 2002, ApJ, 568, 733
\reference{brun02}
Nisini, B. et al. 2002, A\&A, 393, 1035
\reference{nor93}
Noriega-Crespo, A., Garnavich, P.M \& Raga, A.C. 1993, AJ, 106, 1133
\reference{nor02} 
Noriega-Crespo et al. 2002, AAS, 20114202
\reference{nor04}
Noriega-Crespo, A. et al. 2004, ApJS, 154, 352
\reference{alex02a}
Raga, A. C., et al.  2002, A\&A, 395, 647
\reference{alex02b}
Raga, A. C. et al. 2002, ApJ, 565, 29
\reference{alex09}
Raga, A. C.,
Esquivel, A., Vel\'azquez, P. F., Cant\'o, J.,
Haro-Corzo, S., Riera, A., Rodr\'\i guez-Gonz\'alez, A.
2009, ApJ, 707, L6
\reference{alex11} 
Raga, A. C. et al. 2011, arXiv:0184183
\reference{bo89} 
Reipurth, B. 1989 Nature, 340, 44
\reference{bo92}
Reipurth, B., Raga, A. C., Heathcote, S.  1992, ApJ, 392, 145
\reference{bo99} 
Reipurth et al. 1999, A\&A, 352, L83
\reference{rie04}
Rieke, G. H. et al. 2004, ApJS, 154, 25
\reference{luisf08}
Rodr\'\i guez, L. F. et al. 2008, AJ, 136, 1852
\reference{krs93}
Stapelfeldt, K. R. \& Scoville, N. Z. 1993, 408, 239 
\reference{ter99}
Terquem, C., Eisl\"offel, J., Papaloizou, J. C. B.,
Nelson, R. P. 1999, ApJ, 512, L131    
\reference{tob07}
Tobin, J. J., Looney, L. W., Mundy, L. G.,
Kwon, W., Hamidouche, M. 2007, ApJ, 659, 1404
\reference{yang07}
Yang, J. et al. 1997, ApJ, 475, 683
\reference{yba09}
Ybarra, J. E., Lada, E. A. 2009, ApJ, 695, 12
\reference{wer04}
Werner, M. W. et al. 2004, ApJS, 154, 1
\end{references}
\end{document}